\documentclass[prb,twocolumn]{revtex4-2}
\usepackage{amsmath}
\usepackage{color}
\usepackage{graphicx}

\newcommand{\br}{\mathbf{r}}
\newcommand{\bk}{\mathbf{k}}
\begin{document}
	
\title{Dynamics of Conducting Ferroelectric Domain Walls}
\author{Carson Carroll and W. A. Atkinson$^\ast$}
\date{\today}
\affiliation{Department of Physics \& Astronomy, Trent University, Peterborough, Ontario K9L 0G2, Canada}
\email{billatkinson@trentu.ca}
\begin{abstract}
We report on the dynamics of a conducting domain wall under applied dc and ac voltages.  These dynamics are modeled for a thin film that hosts an ideal charged domain wall via a combination of time-dependent Ginzburg-Landau equations for the polarization, the Schr\"odinger equation for the electron gas, and Poisson's equation for the electrostatic potential.  The electron dynamics are treated within a Born-Oppenheimer approximation.  We find that the electron gas introduces an additional degree of freedom, beyond polarization relaxation, that modifies the dynamical response of the domain wall.  While marginally relevant for the dc response, the electron dynamics have a pronounced effect on the film's ac dielectric function.  The dielectric function has an intrinsic contribution, due to the bulk susceptibility of the film, and an extrinsic contribution due to the domain-wall displacement.  The elecron gas affects the dielectric function by changing both the amplitude and phase of the displacement.
\end{abstract}

\maketitle

\section{Introduction}

Ferroelectric devices are strongly influenced by the dynamics of domain walls, which separate regions of different polarization \cite{Damjanovic:1998}. In recent years, attention has shifted towards using the domain walls---rather than the macroscopic polarization---as functional elements of nanoscale devices  \cite{Meier:2022}.  This shift is driven in large part by repeated confirmation of early proposals \cite{Guro:1968,Guro:1970,Krapivin:1970,Vul:1973} that, under the right circumstances, domain walls may be made conducting \cite{Bednyakov:2018}.  Motivated by this, numerous groups have demonstrated that conducting domain walls may form the core of reconfigurable circuits with tunable and nonvolatile properties \cite{Sharma:2017,Ma:2018,Jiang:2018,Sharma:2019,McConville:2020,Xiong:2021,Risch:2022,Wang:2022,Liu:2023}. There is now a push to develop techniques to reproducibly engineer conducting domain walls \cite{Ratzenberger:2024,Bednyakov:2023}, and at a more fundamental level to understand domain wall conduction mechanisms \cite{Tselev:2016,Beccard:2022,Risch:2022,Zahn:2024}.  

In this work, we focus on the motion of conducting domain walls under applied fields.  The dynamics of conventional (insulating) domain walls have been extensively studied in the context of macroscopic polarization switching, for example, in a ferroelectric memory element \cite{Damjanovic:1998}.  Advances in imaging technology have allowed detailed measurements of the dynamics at different stages of the switching process,  including motion of existing domain walls \cite{Merz:1954,McGilly:2015} and nucleation and growth of domain walls at surfaces and in the bulk \cite{Zhang:2020Nucleation,Shur:2021}.  In recent years, theoretical work has tended to focus on microscopic considerations---for example, surface screening \cite{Eliseev:2008} and thermal fluctuations \cite{Boddu:2017,Indergand:2020,Indergand:2021,Bauer:2022,Neumayer:2023,Khachaturyan:2024}---that affect these processes.  

There are reasons to expect that these considerations will be different for conducting domain walls than for insulating walls.  The latter are typically close to electrically neutral, having only a small  bound charge $-\nabla \cdot {\bf P}$  (${\bf P}$ is the ferroelectric polarization).  Conducting domain walls, conversely, form a subset of so-called charged domain walls, for which the bound charge is nonzero.  Under ordinary circumstances, charged domain walls are energetically unstable because they produce large depolarizing electric fields; their stabilization therefore requires that a screening charge---typically oxygen vacancies, itinerant holes, or free electrons---collects at the domain wall \cite{Bednyakov:2018}.  The presence of charges and associated depolarizing fields can have a pronounced effect on the domain-wall morphology \cite{atkinson2022evolution,cornell2023influence,marton2023zigzag,marton2025pyramidal} and nucleation energy \cite{sturman2023effect,Fang:2024}. In the most-studied materials, BaTiO$_3$ and LiNbO$_3$, conducting domain walls have a positive bound charge that is at least partly screened by itinerant electrons.  As a result, the domain walls form conducting channels that  may be manipulated by external fields and strains.   Any theoretical description of their dynamics thus needs to include the dynamics of the itinerant electron gas as well as of the lattice polarization. 

Of particular relevance to this work, Mokr\'y \textit{et al.}\cite{Mokry:2007} and Gureev \textit{et al} \cite{Gureev:2012} obtained an expression for the force per unit area on a charged domain wall subjected to a static external electric field,
\begin{equation}
{\bf f} = \left [ \Delta \Phi  - {\bf \overline E} \cdot \Delta {\bf P} \right] {\bf \hat n} 
+ \sigma_f {\bf \overline E},
\label{eq:Gureev}
\end{equation}
with $\Delta \Phi = \Phi_2-\Phi_1$ the free energy density difference between domains 2 and 1, on either side of the wall, $\sigma_f$ the free charge density at the wall,  ${\bf \hat n}$ the unit normal pointing into region 2, and ${\bf \overline E} = ({\bf E}_2 + {\bf E}_1)/2$ the electric field at the domain wall.  This result is physically appealing:  the first term describes the pressure on the domain wall due to the free energy density and the direct interaction of the electric field with the polarization; the second term describes the force of the electric field on the free charge density.  However, this formula makes a key assumption that the itinerant electron density can be treated as an infinitely thin sheet that is rigidly attached to the domain wall.  One of the main topics explored here is the extent to which the response of the itinerant electron gas modifies the domain wall dynamics. 

Our approach is to solve the time-dependent Landau-Ginzburg-Devonshire (LGD) equations in the presence of a free electron gas that is treated quantum mechanically within the Born-Oppenheimer approximation.  This is a novel approach to the problem, and in part the current work represents a proof-of-principle.  We restrict ourselves to a planar geometry  allows us to separate the response of the electron gas to an applied electric field from that of the lattice polarization.  We can identify three  features that are unique to conducting domain wall dynamics:  (i) on time scales shorter than the characteristic relaxation time of the polarization, domain-wall dynamics are driven by the motion of the electron gas; (ii) at finite bias voltage, the electron density on the domain wall changes as the domain wall moves; and (iii) while the domain wall velocity is still largely determined by the relaxation time, there is a novel fast-switching mechanism driven by depolarizing fields that emerges when the applied bias voltage is sufficiently large.

\begin{figure}
\includegraphics[width=\columnwidth]{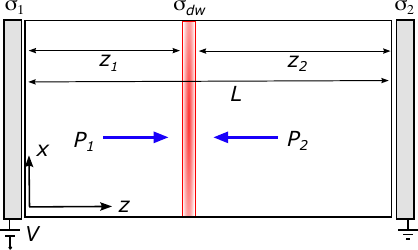}
\caption{Schematic of the model geometry.  Capacitor plates with charge densities $\sigma_1$ and $\sigma_2$ sandwich a ferroelectric thin film of thickness $L$ and infinite extent in the $x$ and $y$ directions.  The plates are held at a fixed voltage difference $V$.  A head-to-head domain wall, a distance $z_1$ from the left surface and $z_2 = L-z_1$ from the right surface, has a net 2D charge density $\sigma_\mathrm{dw}$ and separates domains with polarization $P_1 > 0$ and $P_2 <0$. }
\label{fig:model}
\end{figure}

For concreteness, we imagine a thin-film ferroelectric whose polarization axis lies along the film $c$-axis, and a screening charge made up of free electrons (Fig.~\ref{fig:model}).   This kind of system has, for example, been engineered by compressively straining LaAlO$_3$/SrTiO$_3$ bilayers \cite{Bark:2011}.  There, the thin LaAlO$_3$ cap layer (typically $~\sim 10$ monolayers) acts as a charge reservoir, and donates electrons to the SrTiO$_3$ film with typical densities of $n_\mathrm{2D} \sim 10^{13}$--$10^{14}$~cm$^{-2}$.  In our model, the ferroelectric is sandwiched between capacitor plates that allow direct control of the voltage across the film.  The key assumption is that there is a flat head-to-head domain wall running parallel to the film surface.  While this undoubtedly oversimplifies aspects of the domain-wall structure, we found previously that flat head-to-head structures like the one shown in Fig.~\ref{fig:model} are stable when the density of screening charges is high \cite{cornell2023influence}.

\section{Calculations}
\label{sec:calculations}
We simulate the system shown in Fig.~\ref{fig:model}.  A thin ferroelectric film (thickness $L$ and planar area $A \gg L^2$) hosts a free electron gas with two-dimensional (2D) density $n_\mathrm{2D}$.  The $z$-axis is perpendicular to the film surfaces.  The film is sandwiched between capacitor plates with surface charge densities $\sigma_i$ ($i=1,2$) and held at potentials $V$ (left plate; at $z=0$) and zero (right plate; at $z=L$).  Overall charge neutrality requires that $\sigma_1+\sigma_2 -en_\mathrm{2D} = 0$.  We make the key simplifying assumption that there is translational invariance in the $x$-$y$ plane, such that both the polarization ${\bf P}(z)$ and the three-dimensional electron density $n(z)$ are functions of $z$ only.   With this ansatz, the bound charge density, $-\nabla \cdot {\bf P} = -d P_z / dz$, depends only on the $z$-component of the polarization.  To further simplify the model, we therefore retain only the $z$-component $P\equiv P_z$ in the LGD free energy.  

Figure~\ref{fig:model} shows a domain wall located at $z=z_1$, and the polarization is shown schematically as having values $P_1>0$  and $P_2<0$ to the left and right of the wall, respectively. This generates a positive bound charge along the domain wall which confines electrons to the vicinity of the wall.  The net domain wall charge is $\sigma_\mathrm{dw} = P_1 -P_2 +\sigma_f$; when the electron gas is entirely confined to the domain wall, $\sigma_f = -en_\mathrm{2D}$.  In practice, domain walls are approximately neutral (i.e.\ $|\sigma_\mathrm{dw}| \ll |\sigma_f|$, $|P_1|$, $|P_2|$).  We denote the bulk polarization, in the absence of depolarizing fields, by $P_s$.  Since $P_s$ is the upper-limit value of $|P_1|$ and $|P_2|$,  an electron density of up to $\sim 2P_s/e$ can be accommodated by the domain wall.   We choose a smaller value, $n_\mathrm{2D}=1.4P_s/e$ (the electron charge is $-e$), so that the electrons remain bound to the domain wall even when the bias voltage is nonzero.   This has the consequence that $P_1 = -P_2 \approx  0.7 P_s$ for symmetric configurations.  

The electron dynamics are treated under the Born-Oppenheimer approximation at temperature $T=0$; namely,  the polarization evolves adiabatically on the timescale of the electrons, which therefore remain in equilibrium throughout the simulation.  Self-consistent solutions to the Schr\"odinger and Poisson equations for the electronic eigenstates and electrostatic potential, respectively, are obtained  at each time step (see Appendix~\ref{app:EE}).  This provides an updated electron density $n(z)$ and electrostatic potential $\phi(z)$.

The dynamics of the ferroelectric polarization are modeled via the time-dependent LGD equations,
\begin{equation}
\Gamma  \frac{\partial P}{\partial t} = -\frac{1}{A}  \frac{\delta {\cal F}}{\delta P},
\label{eq:TDLGDE}
\end{equation}
where $\Gamma$ determines the relaxation rate for the polarization and ${\cal F}$ is the total system energy, including contributions from the polarization, the electrons, and the electric fields.  We write
 $\Gamma = |a_1| \tau$, where $a_1$ is the coefficient of the quadratic contribution to the LGD energy; this defines a characteristic relaxation time scale $\tau$.    We show in  Appendix \ref{app:A} that, for a planar geometry, the energy is
\begin{equation}
{\cal F} = {\cal F}_\mathrm{lat} + {\cal F}_\mathrm{el} + {\cal F}_\mathrm{V}
\label{eq:Fparts}
\end{equation}
with 
\begin{eqnarray}
{\cal F}_\mathrm{lat} &=& A \int_0^L  \Phi(P)   dz, \\
{\cal F}_\mathrm{el} &=& \left \langle  -\frac{\hbar^2\nabla^2}{2m^\ast} \right \rangle, \label{eq:F2} \\
{\cal F}_V &=&A  \int_0^L \left [ -\epsilon_b\frac{E^2}{2} + e n \phi - PE \right ] dz, 
\label{eq:F3}
\end{eqnarray}
where 
\begin{equation}
\Phi(P)=g_{11} \left( \frac{dP}{dz} \right )^2 + a_1 P^2 + a_{11} P^4,
\end{equation} 
is the LGD energy density of the polar phase and $D(z) = \epsilon_b E(z) + P(z)$ is the electric displacement. Here, $\epsilon_b$ is the background permittivity \cite{levanyuk2016background}.

We take the boundary conditions $\partial_zP|_0 = \partial_z  P|_L =0$ and $n(0) = n(L)=0$ for the electron density.  The  latter conditions arise naturally from our choice of a hard-wall potential in solving the Schr\"odinger equation.  As discussed in Appendix \ref{app:fourier}, we find it convenient to express $P(z)$ and $n(z)$ as Fourier cosine and sine series, respectively, as these automatically respect the boundary conditions.  It is then straightforward to reformulate  Eq.~(\ref{eq:TDLGDE}) for the Fourier components of $P(z)$ (Appendix \ref{app:TDLGDE}).

\begin{table}
\begin{tabular}{c|c||c|c}
Parameter & Value & Parameter & Value \\
\hline
$g_{11}$ & $2\times 10^{-10}$ Jm$^3$/C$^2$ & $m^\ast$ & $5 m_0$ \\
$a_1$ & $-3\times 10^8$ Jm/C$^2$ & $\epsilon_b$ & $4.5 \epsilon_0$ \\
$a_{11}$ & $1.5 \times 10^9$ Jm$^5$/C$^4$ & $n_\mathrm{2D}$ & $1.4\times 10^{14}$ cm$^{-2}$
\end{tabular}
\caption{Model parameters.}
\label{table1}
\end{table}

Model parameters, given in Table~\ref{table1}, are chosen to reflect typical ferroelectrics.  The LGD parameters yield a saturated polarization $P_s = \sqrt{-a_1/2a_{11}} = 0.32$~C/m$^2$ and correlation length $\xi_0 = \sqrt{-g_{11}/a_1} = 0.8$~nm.  The large electron effective mass, $m^\ast = 5 m_0$ with $m_0$ the bare mass, enhances the 2D density of states by a factor of 5 over its bare value.  This provides a crude way to account for the multiple $t_{2g}$ orbitals that contribute to the conduction bands in typical electron-doped perovskites.  The main consequence of the enhanced mass is to make the domain wall narrower \cite{Gureev:2011,sturman_quantum_2015,cornell2023influence}.

To make sense of the simulations, we employ a model free energy that provides a transparent, albeit simplified, description of a charged domain wall (see Appendix~\ref{app:toy}). The model describes a head-to-head domain wall that hosts a compensating 2D electron gas with electron density $n_\mathrm{2D}$.  To make the model analytically tractable, both the domain wall and electron gas are treated as infinitely thin sheets and the electron gas is assumed to move rigidly with the domain wall.  The equations of motion are then
\begin{eqnarray}
 \frac{\partial P_i}{\partial t} &=& - \frac{1}{Az_i\Gamma}  \frac{\partial {\cal F}}{\partial P_i} 
 =-\frac{1}{ |a_1| \tau} \left[ \frac{d\Phi_i}{dP_i} - E_i \right ] \label{eq:dPdt} \\
 \frac{\partial z_1}{\partial t} &=&  - \frac{\xi_0}{P_s^2} \frac 1{A\Gamma'} \frac{\partial {\cal F}}{\partial z_1}
= -\frac{\xi_0}{ P_s^2 |a_1|  \tau'} \ p\big |_V, \label{eq:dzdt}
\end{eqnarray} 
where ${\cal F}$ is given by Eq.~(\ref{eq:f}), $\Phi_i = a_1 P_i^2 + a_{11} P_i^4$ is the free energy density in region $i = 1,2$, the factor $\xi_0/P_s^2$ is introduced in Eq.~(\ref{eq:dzdt})  to make the units consistent, and $\tau'$ is the characterisic time scale for domain wall motion. Note that $\tau'$ is distinct from $\tau$, which is the characteristic time for the magnitude of $P_i$ to change.  The pressure $p|_V$ in Eq.~(\ref{eq:dzdt}) is the extension of  Eq.~(\ref{eq:Gureev}) to the case of a fixed potential (rather than fixed field) for the planar domain-wall geometry considered in this work:
\begin{equation}
\left . p \right |_{V} = \Phi_2 - \Phi_1 + \sigma_{dw}\left [\frac VL + \frac{z_1-z_2}L \frac{\sigma_{dw}}{2\epsilon_b} \right].
\label{eq:dfdz}
\end{equation}
  A similar expression was recently obtained by Sturman and Podivilov \cite{sturman2023effect}, who noted in passing that it can have the opposite sign to the applied electric field $V/L$, and therefore generate domain wall motion opposite to what is expected based on the sign of $\sigma_{dw}$.  

 \section{Results}
 \subsection{Response to a dc bias voltage}
\begin{figure}
\includegraphics[width=\columnwidth]{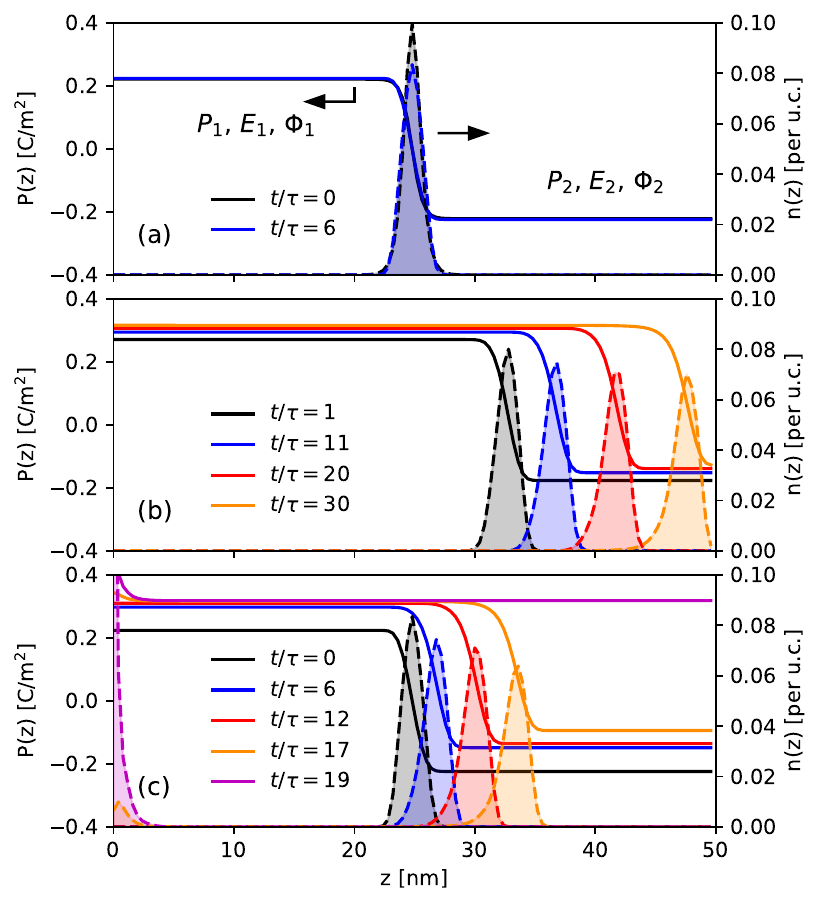}
\caption{Polarization (solid lines) and electron density (dashed lines) at representative times during the  simulations for (a), (b) $V=0$, and (c) $V=1$~V.  In (a) and (b), the initial values of $P(z)$ and $n(z)$  are set by hand such that the domain wall initially sits $z=\frac L2$ and $z=\frac{2L}3$, respectively. In (c), the initial values for $P(z)$ and $n(z)$ are taken from the end of the simulation shown in (a), and the bias voltage is switched on suddenly at $t=0$.  In (b) the domain wall moves to the right without stopping.  In (c), the domain wall first moves to the right and then the electron gas spills over to the interface. 
The electron density is $n_\mathrm{2D} = 1.4 P_s/e$.}
\label{fig:mid1}
\end{figure}

Figure~\ref{fig:mid1} illustrates the domain wall motion for a variety of initial configurations and bias voltages.    Figure~\ref{fig:mid1}(a) shows the evolution of a domain wall that is initially placed at $z=\frac L2$ and is subject to a vanishing bias voltage ($V=0$). We take an initial profile with $P = -0.7P_s \tanh(z/d)$ and $n(z) = dP/dz$, where $d$ is adjusted by hand to reasonably match the domain wall profile obtained by iterating the equations of motion.  As shown in Fig.~\ref{fig:mid1}(a), the domain wall relaxes slightly, but does not move.  This is unsurprising, and is indeed consistent with Eq.~(\ref{eq:dfdz}):
 By symmetry, $P_1=-P_2$, $E_1 = -E_2$, and $\Phi_1=\Phi_2$, such that the pressure obtained from  Eq.~(\ref{eq:dfdz}) vanishes.  
 
\begin{figure}
\includegraphics[width=\columnwidth]{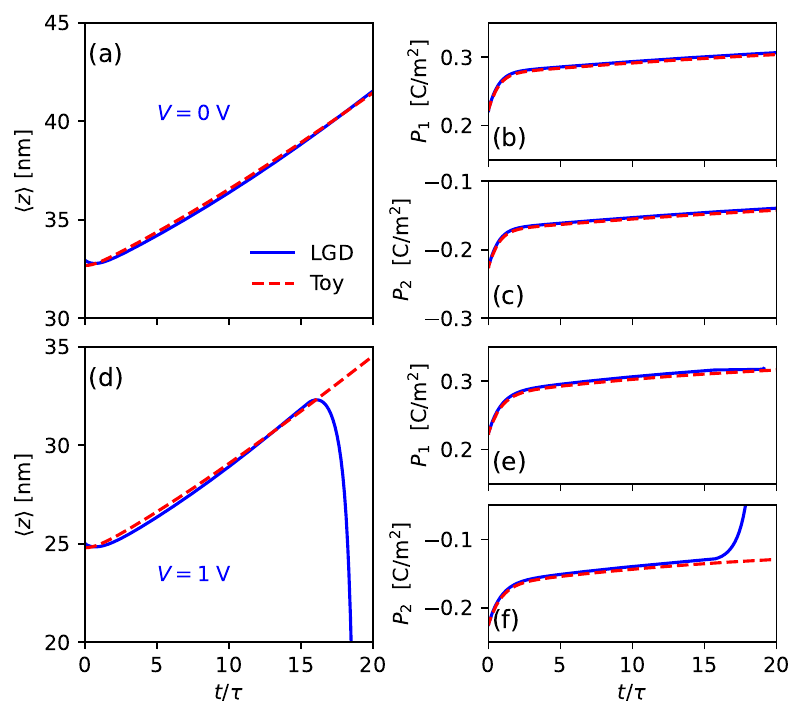}
\caption{Comparison between numerical simulations and the toy model. (a) Domain wall position and (b), (c) polarization to the left and  right of the domain wall for same case as in Fig.~\ref{fig:mid1}(b), namely with initial position $z = \frac {2L}3$ and voltage $V=0$~V.  (d) Domain wall position and (e), (f) polarizations for a domain wall that is initially at $z = \frac L2$ with bias voltage $V=1$~V.   Numerical simulations are obtained by integrating Eq.~(\ref{eq:TDLGDE}), toy model results are obtained from Eqs.~(\ref{eq:dPdt}) and (\ref{eq:dzdt}) with $\tau' = 2\tau$.  The domain wall position is obtained from the center-of-mass of the electron density and coincides with $z_1$ for the toy model.}
\label{fig:LGD-toy}
\end{figure}

Figure~\ref{fig:mid1}(b) shows that when the domain wall starts off-center, it drifts towards the closest surface, even when $V=0$. Such behaviour is naively expected from Eq.~(\ref{eq:dfdz}) because of the term proportional to $z_1-z_2$, which destabilizes the equilibrium point at $z=\frac L2$.  In fact, this term is negligible and the dominant contribution to the pressure comes from the difference $\Phi_2 - \Phi_1$ of the energy densities on either side of the domain wall.  Because of the small film thickness,  the polarizations $P_1$ and $P_2$ depend on the distance between the domain wall and the surfaces.  This is illustrated in Fig.~\ref{fig:LGD-toy}:  As the wall moves rightward [Fig.~\ref{fig:LGD-toy}(a)], $P_1$ grows towards its preferred bulk value [Fig.~\ref{fig:LGD-toy}(b)], which decreases $\Phi_1$; to maintain domain-wall neutrality, $|P_2|$ must shrink by the same amount [Fig.~\ref{fig:LGD-toy}(c)], which increases $\Phi_2$. This produces a large pressure that accelerates the domain wall away from the centre of the film. The flat charged domain walls modeled here are thus unstable in thin films, with the system preferring to form a single-domain phase.  As we show below, however, even a weak pinning force is sufficient to stabilize the domain wall. 

The motion shown in Fig.~\ref{fig:mid1}(b) is a comparatively slow process, being limited by the lattice relaxation rate.  Figure~\ref{fig:LGD-toy}(a) compares this motion to that predicted by Eqs.~(\ref{eq:dPdt}) and (\ref{eq:dzdt}) for the toy model.   We use the same relaxation time $\tau$ as for the LGD simulations, and set  $\tau'=\tau/2$ by trial and error, so that the domain wall velocities obtained from the two approaches are roughly equal when $t > \tau$.  We see that, apart from a discrepancy at small times, the toy model captures the motion reasonably well, and in particular reproduces the values of the polarization accurately [Fig.~\ref{fig:LGD-toy}(b) and (c)].  Note, however, that this depends sensitively on the choice of $\tau'$ and without prior knowledge of the answer, we would not have been able to predict the motion correctly.  

The discrepancy in the initial motion shown in Fig.~\ref{fig:LGD-toy}(a), while small, is interesting.  The toy model predicts that the domain wall moves uniformly in the direction of the pressure, while the time-dependent LGD simulations find a small initial motion in the opposite direction.  As we show below, this effect becomes important when an ac voltage is applied, and we attribute it to the fact that the electron gas is not rigidly attached to the domain wall.

Next, Fig.~\ref{fig:mid1}(c) shows the motion of a domain wall that starts at $z=\frac L2$ and is subjected to a bias voltage  $V=1$~V ($V/L = 200$~kV/cm) that is switched on suddenly at $t=0$.  Our simulations always find that the domain wall has a weak positive charge, such that a positive  voltage  generates a rightward motion.  The motion is compared to the toy model in Fig.~\ref{fig:LGD-toy}(d)-(f) for the same value of $\tau'=\tau/2$ as above.  There is a slight discrepancy for times $t\lesssim \tau$, but otherwise the simulations and the toy model agree well up to $t = 15\tau$.  Inspection of Fig.~\ref{fig:mid1}(c) shows that at $t\sim 15\tau$,  the domain wall reaches a threshold distance from the center at which the electron gas begins to spill over to the positive surface at $z=0$.  Domain wall neutrality is approximately preserved, however, and the bound charge $P_1-P_2$ decreases to compensate for the missing electrons.  In this manner, the domain wall evaporates, leaving behind a single-domain state with the electron gas entirely at the left edge of the film.

The physics behind the electron gas spillover is straightforward:  As the domain wall moves towards regions of lower electrostatic potential, the potential energy of the bound charge decreases while that of the electron gas increases;  eventually, it is energetically favorable for the electron gas to move to the interface.  The surprising aspect of Fig.~\ref{fig:mid1}(c) and Fig.~\ref{fig:LGD-toy} is that this is an extremely fast process relative to the rightward motion of the domain wall, presumably due to the strong depolarizing fields that are generated when the domain wall deviates from neutrality.  We have thus identified a mechanism for fast ferroelectric switching that applies uniquely to the case of conducting domain walls, namely domain wall evaporation.

\begin{figure}
\includegraphics[width=\columnwidth]{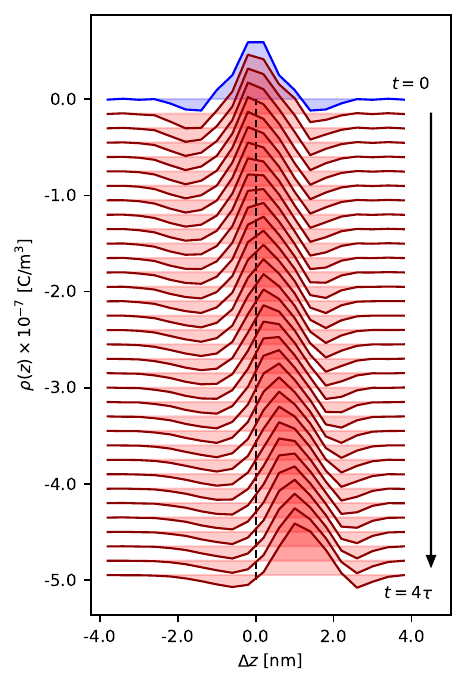}
\caption{Total (bound plus free) charge density as a function of time for the domain wall motion shown in Fig.~\ref{fig:LGD-toy}(d).}
\label{fig:mid3}
\end{figure}

To understand better the discrepancy  between the LGD and toy models for $t\lesssim \tau$, we show an expanded view of the domain wall motion for the  case shown in Fig.~\ref{fig:LGD-toy}(d).  Figure~\ref{fig:mid3} shows snapshots of the total charge density, $\rho(z) = -en_z -\partial P/\partial z$, at  a sequence of times during the initial stages of motion  shown in Fig.~\ref{fig:LGD-toy}(d).  At $t=0$, before the voltage is applied, the charge is distributed symmetrically about the center of the domain wall.   Immediately after the voltage is switched on, the negatively charged electrons move to the left.  This changes the electric field profile in the domain wall, which skews the polarization profile (i.e.\ the shape of the domain wall) such that its centre of mass initially moves leftward.  However, since the domain wall has a net positive charge, there is an overall rightward motion that becomes apparent on times longer than $\tau$.

\subsection{Response to an ac bias voltage}

Having identified that the charged domain wall becomes skewed due to the motion of the electron gas under an applied voltage, we now show that this plays an important role in the ac response of the wall.  We model the domain-wall motion from its initial position at $z = \frac L2$ as a function of time when subjected to an ac bias voltage, $V(t) = V_0 \sin\omega t$ with $V_0 = 0.01$~V ($V/L = 2$~kV/cm).  The voltage is switched on suddenly at $t=0$.  
In the absence of pinning, the  motion has two components:  first, there is an ac response at the driving frequency and, second, the domain wall drifts away from the center of the film.  The latter occurs because, as discussed above, the middle of the film is a point of unstable equilibrium.  To eliminate the drift, we introduce a  pinning potential to the middle of the film, namely we take the LGD parameter $a_1(z)$ to have the form
\begin{equation}
a_1(z) = a_1 \left[1-\alpha a \delta\left(z-\frac L2\right)\right ],
\end{equation}
where $a_1$ is given in Table~\ref{table1}, $\alpha$ is the dimensionless pinning strength, and the lattice constant $a$ is introduced to maintain the correct units.  This model describes a system in which a single layer of weakly- or non-ferroelectric material is $\delta$-doped into the middle of the film.  Pinning occurs when $\alpha > 0$.

\begin{figure}
 \includegraphics[width=\columnwidth]{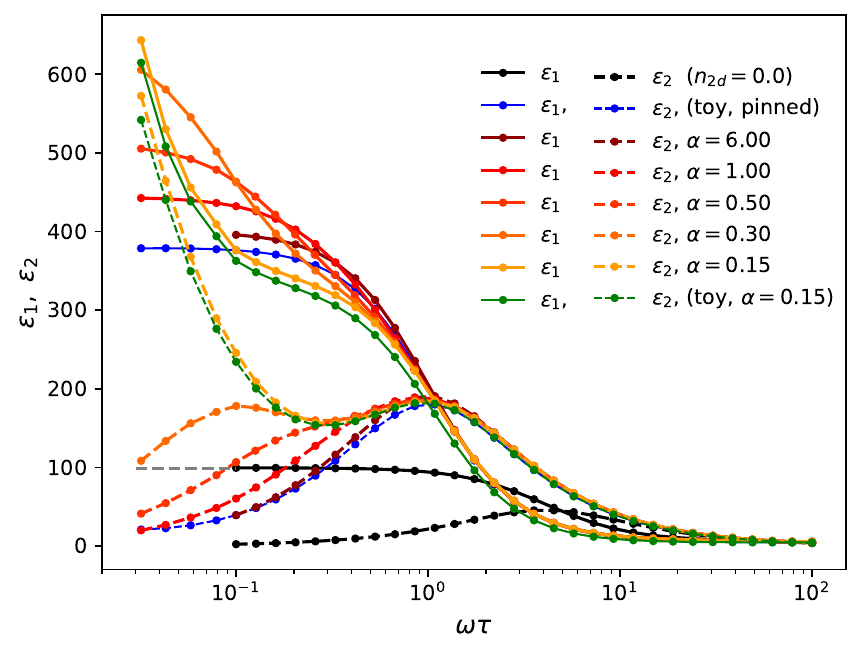}
 \caption{Complex relative permittivity $\varepsilon_r = \varepsilon_1 +i\varepsilon_2$ for a pinned domain wall. The pinning strength ranges from strong ($\alpha=6$) to weak ($\alpha=0.15$).  For comparison, $\varepsilon_r$ is also shown for an insulating ($n_\mathrm{2d}=0$) single-domain thin film, for the toy model with a perfectly pinned domain wall, and for the toy model with a weakly pinned domain wall ($\alpha=0.15$).}
 \label{fig:dielectric}
 \end{figure}

Figure~\ref{fig:dielectric} shows the effective complex dielectric constant, $\varepsilon_r$, obtained by rearranging the capacitance equation ($Q=CV$) to obtain $\varepsilon_r(\omega) = \sigma(\omega)L/\epsilon_0V(\omega)$ in terms of the charge density $\sigma(\omega)$ on the capacitor plates as determined from  Eq.~(\ref{eq:sigma}).  The figure shows the frequency dependence of $\varepsilon_r$ from simulations, along with three reference cases:  a single-domain insulating film ($n_\mathrm{2d} = 0$); the toy model with a perfectly pinned domain wall; and the toy model with a weakly pinned domain wall.  For the single-domain case, the dielectric response is that of the bulk material, i.e.~$\epsilon_0 \varepsilon_r = \epsilon_b + |a_1|^{-1}/(4-i\omega\tau)^{-1}$.

The numerical simulations show that domain-wall pinning has a substantial effect on the dielectric response at low frequencies, $\omega\tau \lesssim 1$.  In the strongly pinned case, the imaginary component $\varepsilon_2$ has a single peak at $\omega\tau\approx 1$. As the pinning strength $\alpha$ is reduced, however, a shoulder emerges on the low-frequency side of the peak.  This shoulder develops into a distinct peak in the weak-pinning limit, at a  frequency that decreases with decreasing $\alpha$.  Pinning also affects the real component $\varepsilon_1$ of the dielectric function.  Relative to the perfectly pinned case, $\varepsilon_1$ is progressively enhanced at low frequencies with decreasing $\alpha$, but actually decreases at intermediate frequencies.  

To gain insight into our numerical simulations, we developed an approximate expression for the dielectric function based on the toy model (Appendix~\ref{sec:dielectric}),  
\begin{eqnarray}
\epsilon_0\varepsilon_r &\approx & \epsilon_b +  \frac{1}{\chi_0^{-1} -i\omega \tau |a_1|} \left[ 1 + \frac{\Delta z_1(\omega) \sigma_\mathrm{dw}^0}{ V(\omega) \epsilon_b} \right ],
\label{eq:effective}
\end{eqnarray}
where $\sigma_\mathrm{dw}^0 = 2P_0-en_\mathrm{2d}$ is the equilibrium domain-wall charge density, $P_0$ is the equilibrium polarization away from the domain wall,  $\Delta z_1 = z_1 - \frac L2$ is the displacement of the domain wall from equilibrium in the middle of the film, and $\chi_0^{-1} = 2a_1 + 12 a_{11}P_0^2$. Equation~(\ref{eq:effective}) contains two terms:  the first gives the intrinsic contribution and is the dielectric function in the perfectly-pinned limit, while the second gives the extrinsic contribution due to the domain-wall motion. The two toy-model curves in Fig.~\ref{fig:dielectric} are generated from Eq.~(\ref{eq:effective}) with $\Delta z_1=0$ (pinned) and $\Delta z_1$ taken from the numerical simulations (toy, $\alpha=0.15$).  In general,  Eq.~(\ref{eq:effective}) reproduces the numerical simulations very well provided that $\Delta z_1(t)$ is accurate.  Importantly, numerical integration of the toy-model equations of motion give results for $\Delta z_1(t)$ that  underestimate both the amplitude and phase shift of the oscillation (see below).  For this reason, the values of $\Delta z_1(\omega)$ used to make the toy-model plots in Fig.~\ref{fig:dielectric} are taken from the full time-dependent LGD simulations.

Equation~(\ref{eq:effective})  shows that the profound changes in $\varepsilon_r(\omega)$ that occur as pinning is reduced are due to the domain wall motion. We can rule out any direct role for internal domain-wall dynamics (for example breathing modes associated with changes in the domain wall width) because the toy model from which Eq.~(\ref{eq:effective}) is derived  (i) matches the numerical data closely and (ii) does not contain that physics.  The key result from Eq.~(\ref{eq:effective}) is that the extrinsic contribution to the dielectric function depends on the complex ratio $\Delta z_1(\omega)/V(\omega)$.  From this, we establish that domain-wall motion that is in-phase (out-of-phase) with the applied voltage will enhance (reduce) $\varepsilon_r$.   

\begin{figure}
\includegraphics[width=\columnwidth]{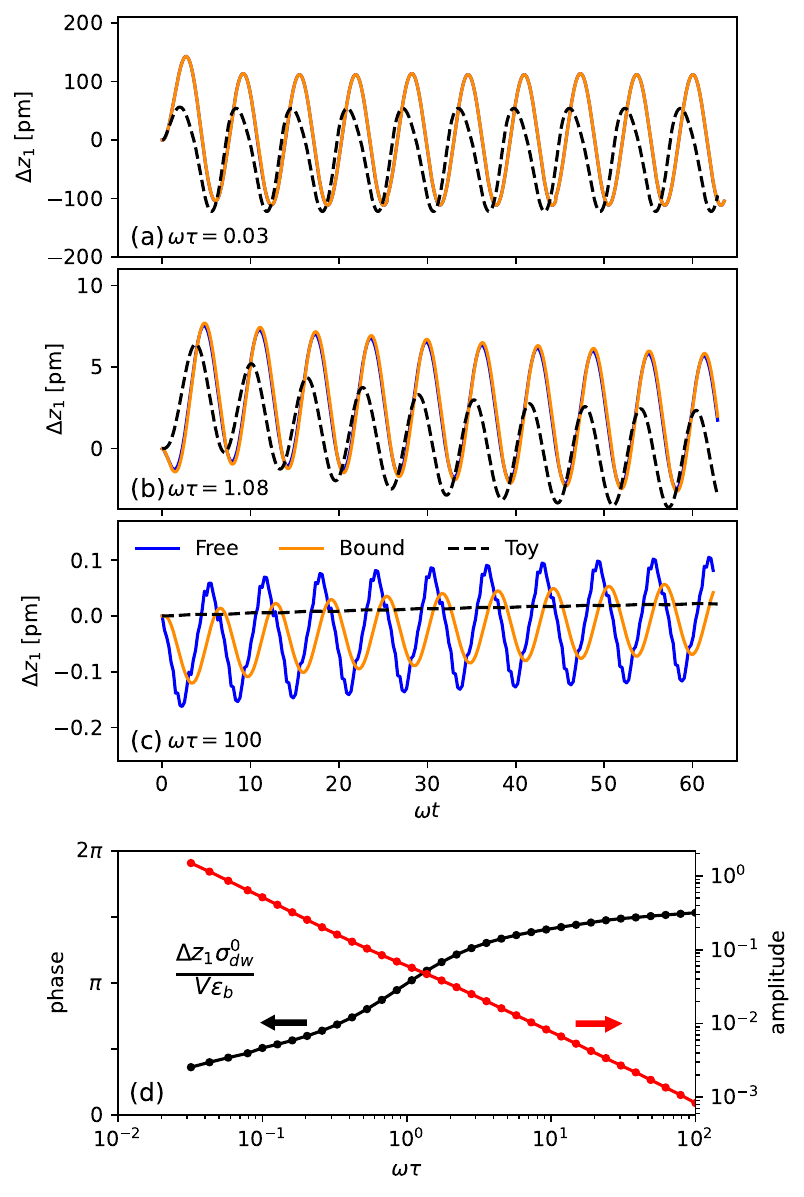}
\caption{Domain-wall motion under an ac voltage $V(t) = V_0 \sin \omega t$ with $V_0 = 0.01$~V and for a weak pinning potential with $\alpha=0.15$. Domain-wall displacements as measured by the free and bound charge densities are shown for (a) $\omega\tau = 0.03$, (b)  $\omega\tau = 1.1$, and (c) $\omega\tau = 100$. In (a), a sine curve qualitatively representing the ac potential is included to highlight the phase difference between the applied voltage and domain-wall position.  The complex phase and amplitude of the term $\Delta z_1(\omega) \sigma_\mathrm{dw}^0/V(\omega)\epsilon_b$ are shown in (d).  This term describes the correction to the dielectric function in Eq.~(\ref{eq:effective}) due to domain-wall motion.  }
\label{fig:osc}
\end{figure}

The domain-wall position  may be obtained from the center of mass of either the free electron density $n(z)$ or the bound charge density $-\nabla\cdot {\bf P}(z)$, and both are plotted in Fig.~\ref{fig:osc} for the weakly pinned case.  At low driving frequencies, $\omega \tau \lesssim 1$, the free and bound charge densities move in phase and with the same amplitude; indeed, the two trajectories are  indistinguishable in Fig.~\ref{fig:osc}(a) and nearly so in Fig.~\ref{fig:osc}(b).  The trajectories are, however, noticeably different for $\omega\tau \gg 1$ [Fig.~\ref{fig:osc}(c)], and in Eq.~(\ref{eq:effective}), we define $\Delta z_1(t)$ from the bound charge density.   
 
One conclusion from Fig.~\ref{fig:osc} is that the key assumption of the toy model that the electron gas is rigidly attached to the domain wall is approximately upheld at the lowest frequencies. However, the toy model and numerical simulations differ in important ways. To illustrate this, we numerically integrate the toy-model equations of motion, which are modified to include the pinning potential, and plot the results in Fig.~\ref{fig:osc}.
%
It is apparent that the toy model underestimates both the amplitude and phase of the domain wall oscillations, especially at high frequencies.  
At all frequencies, the toy-model predicts that $\Delta z_1$ lags the applied voltage by a phase of $\frac \pi 2$, as one would expect for a massless domain wall, while the numerical simulations obtain a frequency-dependent phase that ranges from $\sim \frac \pi 4$ to $\sim \frac{3\pi}4$ [Fig.~\ref{fig:osc}(d)]. In Fig.~\ref{fig:osc}(b) and (c), it is striking that the domain wall initially moves opposite to the applied electric field, despite the wall having a net positive charge.  This physics is entirely missed by the toy model.  On the other hand, both calculations find a power-law frequency-dependence for the amplitude, although the toy-model amplitude decays faster. The key point of this discussion is that although we can rule out direct contributions of internal domain-wall dynamics to $\varepsilon_r(\omega)$, it appears that these make key indirect contributions through the amplitude and phase of $\Delta z_1(\omega)$. 

We end this section with a remark about the differences between the single-domain and strongly pinned ($\alpha=6$) cases.  Domain-wall dynamics are absent in both cases, such that the effective dielectric response is entirely intrinsic (namely, there is no direct contribution from domain walls).  It therefore appears surprising that the low-$\omega$ dielectric constant is 4 times bigger in the domain-wall case.   This is because the equilibrium polarization $P_0$ in the presence of the domain wall is less than the bulk polarization $P_s$ which is obtained in the single-domain case.  Consequently, the susceptibility $\chi_0$ is different from that of the bulk. The key point is that the constraint of (approximate) domain-wall neutrality, $2P_0 -en_\mathrm{2d}\approx 0$, that is implicit in the Landau free energy affects the intrinsic response whenever conducting domain walls are present.  Importantly, $\chi_0$ decreases with increasing electron density, and in the limit $en_\mathrm{2d} = 2P_s$ (for which, $P_0=P_s$) the enhancement vanishes.

\section{Discussion}
The ac responses of insulating and weakly conducting ferroelectrics are a subject of renewed interest \cite{Chu:2014Kinetics,Liu:2017,Wu:2017low,Gurung:2025extrinsic}.   It is well-known that domain walls enhance the effective bulk dielectric constant, and there are conflicting reports as to whether the enhancement comes from domain-wall motion or from internal degrees of freedom that include, for example, breathing modes associated with changes in the domain-wall width. Simulations show that at low electron densities, and for film geometries like those studied here, the domains have a lamellar structure and are nearly perpendicular to the film surfaces \cite{cornell2023influence}.  Because the  walls run parallel to the applied field, the effective force on the  wall---as given by Eq.~(\ref{eq:Gureev})---comes entirely from the action of the field on the polarization, while the force on the electron gas creates a current along the wall. At high electron densities (the case considered here), domain walls tend to lie parallel to the film surfaces and in this case the wall dynamics depend on the force on both the electron gas and the polarization.

In reality, conducting domain walls are usually tilted with respect to the film surface.  Earlier work showed that conducting domain walls tilt relative to the horizontal structure, either forming zigzag or tilted lamellar patterns \cite{cornell2023influence}.  In this case, the electronic motion should be more complicated, with flow along the domain walls as well as motion perpendicular to them.  Furthermore, for thin films both the bound and free charges spread much farther from the centre of the domain wall in the zigzag phase than one would expect from simple planar models.  It is likely that, while the main ideas presented here should still be relevant, the electronic contributions to the domain wall motion will be more complicated when the domain walls deviate from a simple geometry.

Finally, we comment on the validity of the Born-Oppenheimer approximation, as it is applied here. Importantly, we have presumed that there is a single chemical potential throughout the film, and that the electron gas remains in thermodynamic equilibrium throughout the domain wall motion.  Thus, electrons may tunnel between the film surfaces and the domain wall instantly on the time scale of the domain-wall motion.  This is reasonable in ultrathin films, but is problematic in thick films where the time for the electronic subsystem to equilibrate may not be much different from the lattice relaxation time.

\section{Conclusions}
We have studied the motion of an idealized, flat, conducting domain wall under an applied voltage by solving, simultaneously, time-dependent Ginzburg-Landau equations for the polarization and the Schr\"odinger equation for the electronic bands. We find that on timescales less than the characteristic relaxation time $\tau$, the domain-wall dynamics are noticeably affected by the fast dynamics of the electron gas.  While marginally relevant for dc applied fields, the electronic degrees of freedom strongly modify the ac response.  We showed that this modification is indirect:  the effective dielectric function of a film hosting a domain wall depends on both the bulk susceptibility (the intrinsic contribution) and the domain wall displacement (the extrinsic contribution), with the latter being strongly affected by the electron gas.

\appendix
\section{Free Energy}
\subsection{Electron Energy}
\label{app:EE}
Under the Born-Oppenheimer approximation, we assume that the electrons are in equilibrium at each timestep.  Setting the variation of ${\cal F}$ to zero subject to the constraint that the eigenstates are normalized gives the Schr\"odinger equation for the single-electron eigenstates,
\begin{equation}
\left[ \frac{-\hbar^2}{2m^\ast} \nabla^2 - e\phi(z) \right ] \Psi = {\cal E} \Psi
\end{equation}
Translational invariance in the $x$-$y$ plane implies that the eigenstates  are separable, with wavefunctions and eigenenergies
\begin{eqnarray}
\Psi_{n\bk}(\br) &=&\frac{1}{\sqrt{A}} \psi_n(z) e^{i(k_x x  + k_y y)}, \\
{\cal E}_{n\bk} &=& \epsilon_n + \frac{\hbar^2(k_x^2+k_y^2)}{2m^\ast},
\end{eqnarray}
 and
\begin{equation}
\epsilon_n \psi_n(z) = \left[ -\frac{\hbar^2}{2m^\ast}\frac{\partial^2}{\partial z^2} - e\phi(z) \right]\psi_n(z).
\label{eq:SE}
\end{equation}
Equation~(\ref{eq:SE}) is solved numerically on a grid.  From these solutions the electron density is obtained at $T=0$ from
\begin{eqnarray}
n(z) 
&=& \frac{m^\ast}{\pi\hbar^2} \sum_{\epsilon_n < \mu} (\mu-\epsilon_n)|\psi_n(z)|^2.
\label{eq:nz}
\end{eqnarray}
The chemical potential $\mu$ is determined at each timestep by the constraint that the 2D electron density,  
\begin{equation}
n_\mathrm{2D} = \int_0^L n(z) dz,
\label{eq:n2D}
\end{equation}
is fixed.  Then, the electronic energy is
\begin{eqnarray}
\frac{{\cal F}_\mathrm{el}}{A} &=& \frac{m^\ast}{2\pi \hbar^2} \sum_{\epsilon_n < \mu} \left( \mu
^2 - \epsilon_n^2 \right ). 
\label{eq:Fel}
\end{eqnarray}

\subsection{Legendre Transformation to Finite Voltage}
\label{app:A}
If the charge on the capacitor plates is fixed, then the total free energy is 
\begin{equation}
{\cal G} = A\int_0^L \left [ \Phi(P) +  \frac{\epsilon_b}2 E(z)^2 \right ]dz + \langle H_0 \rangle
\label{eq:G}
\end{equation}
where $A$ is the cross-sectional area of the device, 
\begin{equation}
\Phi(z) = a_1 P(z)^2  + a_{11} P(z)^4 + g_{11} \left( \frac{\partial P}{\partial z} \right )^2,
\end{equation}
is the free energy density for the polarization, 
 $E(z)$ is the total electric field, satisfying
\begin{equation}
\epsilon_b \frac{\partial E}{\partial z} = -en(z) - \frac{\partial P}{\partial z},
\end{equation}
and 
\begin{equation}
\hat H_0 = -\frac{\hbar^2\nabla^2}{2m^\ast},
\end{equation}
is the kinetic energy operator for the electron gas.  In practice, experiments are carried out at fixed voltage, and the appropriate free energy is
\begin{equation}
{\cal F} = {\cal G} - \oint_S \sigma \phi da
\end{equation}
where $\sigma$ and $\phi$ are the surface charge density and potential, respectively, on the surface bounding the ferroelectric.  For the capacitor geometry, this integral is over the top and bottom capacitor plates.  By construction, the potentials on the left and right plates are $\phi(0) = V$ and $\phi(L)=0$, respectively.  Taking the surface charge density on the left plate to be $\sigma_1=Q_1/A$, we obtain
\begin{equation}
{\cal F} = A\int_0^L \left [ \Phi(P) +  \frac{\epsilon_b}2 E(z)^2 \right ]dz + \langle H_0 \rangle - Q_1 V 
\label{eq:F}
\end{equation}
with the constraint $\int_0^L E dz = V$.  

The free energy can be cast into the form (\ref{eq:Fparts}), with
\begin{equation}
{\cal F}_V = A \int_0^L \left[ \frac{\epsilon_b}{2} E(z)^2 - \frac{P^2}{2\epsilon_b}  
+ e n(z) \phi(z) \right ] - Q_1 V.
\end{equation}
Writing $en(z) = -d D/dz$, with $D = \epsilon_b E + P$, integrating the final term in the integrand by parts, and noting that $\sigma_1 = D(0)$, we obtain Eq.~(\ref{eq:F3}).

\subsection{Lattice and Field Energies}
\label{app:fourier}
For the geometry shown in Fig.~\ref{fig:model}, it is simplest to work in a Fourier representation.
We adopt Neumann boundary conditions for the polarization
\begin{equation}
\left . \frac{dP}{dz}\right |_{z=0} = \left . \frac{dP}{dz}\right |_{z=L} = 0,
\end{equation}
and the hard-wall boundary conditions on the Schr\"odinger equation ensure that the electron density satisfies,
\begin{equation}
n(0) = n(L) = 0.
\end{equation}
With these conditions, it is simplest to express 
\begin{equation}
P(z) = \sum_{n=0}^N p_n c_n(z), 
\end{equation}
where 
\begin{equation}
c_n(z) = \left \{ \begin{array}{ll} \frac 1{\sqrt{L}}, & n=0 \\  \sqrt{\frac 2L} \cos(k_n z), & n \geq 1
\end{array}
\right .
\end{equation}
and $k_n = n\pi/L$ with $N$ the number of $k$-points in the Fourier series.  We also treat the ferroelectric as an infinite square well, such that the electron wavefunctions vanish at the boundary.  It follows that the electronic charge density can be written as a Fourier sine series,
\begin{equation}
-en(z) = \sum_{n=1}^N q_n s_n(z)
\end{equation}
with $s_n(z) = \sqrt{\frac 2L} \sin(k_nz)$.  In practice, we solve the Schr\"odinger equation on a real-space grid to find the electron density, and then obtain the Fourier coefficients by integration: $q_n = -e \int_0^L n(z) s_n(z) dz$.   From this, we obtain the polarization free energy,
\begin{equation}
\frac{{\cal F}_\mathrm{lat} }{A} = \sum_{j=0}^N \left( g_{11}k_j^2 + a_1 + \frac{\delta_{j\neq 0}}{2\epsilon_b} \right) p_j^2 
 + a_{11} \int_0^L P(z)^4 dz
 \label{eq:Flat}
\end{equation}

To obtain the field energy, we solve Gauss' law  subject to the boundary conditions $\phi(0) = V$, $\phi(L) = 0$ to give
\begin{equation}
\phi(z) = V\left [ 1-\frac zL \right ] + \sum_{n=1}^N \frac{q_n + p_n k_n}{\epsilon_b k_n^2} s_n(z),
\label{eq:phi}
\end{equation}
Then, the electric displacement is\
\begin{equation}
D(z) = \epsilon_b \frac VL + p_0 c_0(z) - \sum_{j=1}^N \frac{q_j}{k_j} c_j(z).
\end{equation}
Note that $D(z)$ does, in fact, depend on the $j=0$ Fourier component of the polarization. It follows directly that
\begin{equation}
\frac{{\cal F}_V}{A} = -\frac{1}{2\epsilon_b} \left [ \left( \epsilon_b \frac V{\sqrt{L}} + p_0 \right )^2 - \sum_{j=1}^N \frac{q_j^2}{k_j^2} \right ].
\label{eq:FV}
\end{equation}

\subsection{Time-dependent Landau-Ginzburg-Devonshire Equations}
\label{app:TDLGDE}
Taking the variations of Eqs.~(\ref{eq:Fel}), (\ref{eq:Flat}), and (\ref{eq:FV}) with respect to $q_j$ and $p_j$ and combining the results, yields 
\begin{eqnarray}
\frac{\partial F}{\partial q_j} &=& 0 \\
\frac{\partial F}{\partial p_j} &=& 2 ( g_{11}k_j^2 + a_1 )p_j + \delta_{j\neq 0} \frac{q_j + p_j k_j}{\epsilon_b k_j} \nonumber \\
&& + 4 a_{11} \int_0^L P(z)^3 c_j(z) dz - \frac{V}{\sqrt{L}} \delta_{j,0}.
\end{eqnarray}
In this expression, $\delta_{j\neq 0} \equiv 1 - \delta_{j,0}$, with $\delta_{j,0}$ the Kronecker delta function.
That the total energy has no linear dependence on $q_j$ is a consequence of the Born-Oppenheimer approximation, which assumes the electronic energy is minimized at each timestep.

These results are straightforward to obtain, with the caveat that $\delta {\cal F}_\mathrm{el}$ requires a few intermediate steps; namely, we need to establish how both the chemical potential $\mu$ and energy eigenvalues $\epsilon_n$ respond to changes in $q_j$ and $p_j$.    Applying first order perturbation theory to Eq.~(\ref{eq:SE}), we obtain the shift in the energy eigenvalues in response to a change in the potential,
\begin{eqnarray}
\delta \epsilon_n &=& -e\langle \psi_n | \delta \phi |\psi_n\rangle = -e\int_0^L \delta\phi(z) |\psi_n(z)|^2 dz \nonumber \\
&=& \frac{1}{2\epsilon_b} \sum_{j=1}^\infty \frac{\delta q_j + k_j \delta p_j}{k_j^2} Q_{jn}
\end{eqnarray}
with 
\[
Q_{jn} = -2e \int_0^L  s_j(z) |\psi_n^{(0)}(z)|^2 dz.
\]
Equation (\ref{eq:phi}) has been used to relate $\delta\phi$ to $\delta p_j$ and $\delta q_j$. 

Next, we obtain $\delta \mu$ from Eq.~(\ref{eq:n2D}).  We set $\delta n_\mathrm{2D} =\frac{m}{\pi \hbar^2} \sum_{\epsilon_n < \mu} (\delta \mu - \delta \epsilon_n) = 0$, from which
\begin{eqnarray}
\delta \mu &=& \frac{1}{2\epsilon_b} \sum_{j=1}^\infty \frac{\delta q_j + k_j \delta p_j}{k_j^2}\frac{1}{n_\mathrm{occ} } \sum_{\epsilon_n < \mu} Q_{jn} 
\end{eqnarray}
where $n_\mathrm{occ}$ is the number of occupied bands.
Then,
\begin{eqnarray}
\delta {\cal F}_\mathrm{el} &=& \frac{m}{\pi \hbar^2} \sum_{n=1}^{n_\mathrm{occ}} (\mu \delta \mu - \epsilon_n \delta \epsilon_n ) \nonumber \\
&=& \frac{m}{2\pi \hbar^2\epsilon_b}  \sum_{j=1}^\infty \frac{\delta q_j + k_j \delta p_j}{k_j^2} \sum_{n=1}^{n_\mathrm{occ}}
\left (\mu   -  \epsilon_n \right) Q_{jn} \nonumber \\
&=& \frac{1}{\epsilon_b} \sum_{j=1}^\infty \frac{\delta q_j + k_j \delta p_j}{k_j^2}  q_j.
\end{eqnarray}

\section{Toy Model}
\label{app:toy}
  As before, the capacitor plates at $z=0$ and $z=L$ carry surface charge densities $\sigma_1$ and $\sigma_2$, respectively, while the domain wall carries a net charge density $\sigma_\mathrm{dw} = P_1 - P_2 -e n_\mathrm{2D}$.  To preserve system neutrality, we write
\begin{equation}
\sigma_1 = \sigma + \frac{en_\mathrm{2D}}2; \quad
\sigma_2 = -\sigma + \frac{en_\mathrm{2D}}2,
\label{eq:s1s2}
\end{equation}
where $\sigma$ is the charge density transferred between the capacitor plates by the voltage $V$.   We denote the regions to the left and right of the domain wall as region 1 and region 2, respectively.  The  energy density in region $j$ is $\Phi_j = a_1 P_j^2 + a_{11} P_j^4$. 

\subsection{Pressure at fixed electric field}
To begin, we consider the case for which the charge transfer $\sigma$ is held fixed during the domain wall motion.
Solving Gauss' Law and making use of Eq.~(\ref{eq:s1s2}), gives the electric fields to the left and right,
\begin{eqnarray}
\epsilon_b E_1 &=& \sigma + \frac{en_\mathrm{2D}}2 - P_1, \\
\epsilon_b E_2 &=&\sigma - \frac{en_\mathrm{2D}}2 - P_2.
\end{eqnarray}
These expressions are independent of $z_1$ and $z_2$, and the electric fields are consequently invariant under a virtual displacement of the domain wall.

The appropriate LGD free energy, ${\cal G}[P_1, P_2, z_1, \sigma ]$, is 
\begin{eqnarray}
\frac{{\cal G}}{A} &=& \int_0^L \left[ \Phi[P] +  \frac{\epsilon_b}2 E^2 \right ] dz \nonumber \\
&=& z_1 \Phi_1 + z_2 \Phi_2 + \frac{\epsilon_b}2 ( z_1 E_1^2 + z_2 E_2^2 ).
\end{eqnarray}
Under a virtual displacement $\delta z_1= -\delta z_2$, we obtain an expression for the pressure on the domain wall at constant electric field,
\begin{equation}
p \big |_{E} = - \frac{1}{A}\frac{\partial {\cal G}}{\partial z_1} = \Phi_2 - \Phi_1 + \frac{\epsilon_b}2 (E_2^2-E_1^2).
\end{equation}
Using Gauss' law, $\epsilon_b(E_2 - E_1) = \sigma_\mathrm{dw}$, we arrive at Eq.~(\ref{eq:Gureev}) for the case ${\bf E} \| {\bf P} \| {\bf \hat n}$,
\begin{equation}
p \big |_{E} = {\bf f\cdot \hat n} = \Phi_2 - \Phi_1 + \frac{E_1+E_2}2 \sigma_\mathrm{dw},
\label{eq:Mokry}
\end{equation}

\subsection{Pressure at fixed voltage}
At constant voltage $V$  (see Appendix \ref{app:A}), the appropriate free energy is  ${\cal F}[P_1, P_2, z_1, V]$, where
\begin{equation}
\frac{{\cal F} }{A} =z_1 \Phi_1 + z_2 \Phi_2 + \frac{\epsilon_b}2 \left[ z_1 E_1^2 + z_2 E_2^2 \right ] 
- \left( \sigma -\frac{en_\mathrm{2D}}2 \right ) V.
\label{eq:f}
 \end{equation}
In this case, both $E_1$ and $E_2$ are implicit functions of $z_1$ through the surface charge density $\sigma$.  
Solving Gauss' law subject to the constraint 
 \begin{equation}
 V = \int_0^L E dz =  E_1 z_1 + E_2 z_2
 \end{equation}
 yields 
 \begin{eqnarray}
 E_1 &=& \frac VL - \frac {z_2}L \frac{\sigma_\mathrm{dw}}{\epsilon_b}, \label{eq:E1} \\
E_2 &=& \frac VL + \frac {z_1}L \frac{\sigma_\mathrm{dw}}{\epsilon_b}  \label{eq:E2}  
\end{eqnarray}
and
\begin{equation}
\sigma = \epsilon_b \frac{E_1 + E_2}2 + \frac{P_1+P_2}2.
\label{eq:sigma}
\end{equation}
Then,
\begin{eqnarray}
\left . p \right |_{V} &=& -\frac 1A \frac{\partial {\cal F}}{\partial z_1} \nonumber \\
&=&  \Phi_2 - \Phi_1 + \frac{\epsilon_b}2 (E_2^2-E_1^2) \nonumber \\
&&- \epsilon_b \left [ \left( z_1 E_1 -\frac V2 \right ) \frac{\partial E_1}{\partial z_1} + \left( z_2 E_2 -\frac V2 \right ) \frac{\partial E_2}{\partial z_1} \right ]  \nonumber \\
&=& \Phi_2 - \Phi_1 + \sigma_{dw}\left [\frac VL + \frac{z_1-z_2}L \frac{\sigma_{dw}}{2\epsilon_b} \right].
\end{eqnarray}

\subsection{Effective dielectric function}
\label{sec:dielectric}
 Equations~(\ref{eq:E1}), (\ref{eq:E2}), and (\ref{eq:sigma}) can be combined to obtain an expression for the effective dielectric function for the ferroelectric,
\begin{equation}
\epsilon_0\varepsilon_r = \frac{\sigma}{ V/L} 
= \epsilon_b + \frac{(z_1-z_2)\sigma_\mathrm{dw}}{V} + \frac{L(P_1+P_2)}{2V}.
\end{equation}
These terms represent the background, direct domain-wall motion, and induced polarization contributions.  Of these three, the third is the largest by far.
We proceed further by writing $P_1 = P_0 + p_1$, $P_2 = -P_0 + p_2$, where $P_0$ is the equilibrium polarization, and writing the linearized equations of motion for $p_1$ and $p_2$ [c.f.\ Eq.~(\ref{eq:dPdt})] as
\begin{eqnarray}
\dot p_1 &=& \frac{-1}{|a_1|\tau}\left [ \chi_0^{-1} p_1 -  \frac VL - \frac{\sigma^0_\mathrm{dw} \Delta z_1 }{L\epsilon_b} + \frac{(p_1-p_2)}{2\epsilon_b}  \right]\\
\dot p_2 &=& \frac{-1}{|a_1|\tau}\left [ \chi_0^{-1} p_2 -  \frac VL - \frac{\sigma^0_\mathrm{dw} \Delta z_1 }{L\epsilon_b} - \frac{(p_1-p_2)}{2\epsilon_b}  \right]
\end{eqnarray}
where $\sigma_\mathrm{dw}^0 = 2P_0-en_\mathrm{2d}$ is the equilibrium domain-wall charge density,  $\Delta z_1 = z_1 - \frac L2$ is the displacement of the domain wall from equilibrium in the middle of the film, and $\chi_0^{-1} = 2a_1 + 12 a_{11}P_0^2$.  Solving these at frequency $\omega$ gives
\begin{eqnarray}
\epsilon_0\epsilon_r &\approx & \epsilon_b + \frac L{V(\omega)} \frac{p_1(\omega) +p_2(\omega)}{2} \nonumber \\
&=& \epsilon_b +  \frac{1}{\chi_0^{-1} -i\omega \tau |a_1|} \left[ 1 + \frac{\Delta z_1(\omega) \sigma_\mathrm{dw}^0}{ V(\omega) \epsilon_b} \right ].
\end{eqnarray}
When $\Delta z_1(\omega) = 0$, we obtain the dielectric function of the perfectly-pinned domain wall.  The dielectric function is enhanced  (reduced) by domain-wall motion that is in-phase (out-of-phase) with the driving potential.

\bibliography{CDW_bib}

\end{document}